\documentclass[11pt,twoside]{article}

\usepackage{asp2006}
\usepackage{epsf}
\usepackage{psfig}
\usepackage{lscape}
\usepackage{graphicx}

\begin{document}
\title{The luminosity-size relation of galaxies to $z=1$?}   
\author{Ewan Cameron\altaffilmark{1}}  
\altaffiltext{1}{Visiting scholar at the University of St Andrews}
\affil{Research School of Astronomy and Astrophysics, Mount Stromlo Observatory, Cotter Road, Weston Creek, A.C.T., 2611, Australia}    
\author{Simon P. Driver}   
\affil{SUPA\altaffilmark{2}, School of Physics and Astronomy, University of St Andrews, North Haugh, St Andrews, KY16 9SS, Scotland}    
\altaffiltext{2}{Scottish Universities Physics Alliance}

\begin{abstract} 
We use the Hubble Ultra Deep Field (UDF) to study the galaxy luminosity-size ($M$-$R_e$) distribution.  With a careful analysis of selection effects due to both detection completeness and measurement reliability we identify bias-free regions in the $M$-$R_e$ plane for a series of volume-limited samples.  We also investigate the colour-log($n$) distribution of these galaxies and further subdivide our data by structural type to separately study compact and diffuse objects.  By comparison to the nearby Millennium Galaxy Catalogue, we present tenative evidence for evolution of diffuse, disk-like galaxies with redshift---both in mean surface brightness and the slope of the $M$-$R_e$ relation.  In contrast we find no evidence of structural evolution in the compact galaxy $M$-$R_e$ relation over this redshift range, although there is a suggestion of colour evolution.  We also highlight the importance of considering surface brightness dependent measurement biases in addition to incompleteness biases.  In particular, the increasing, systematic under-estimation of Kron fluxes towards low surface brightnesses may cause diffuse, yet luminous, systems to be mistaken for faint, compact objects.
\end{abstract}

\vspace{-1.0cm}
\section{Introduction}
Traditionally studies of galaxy evolution for the entire galaxy population have been pursued by looking for variations in the monovariate luminosity function, either for the whole population (e.g., CFRS; \citealt{lil95}), or subdivided by morphological type or colour cuts (VVDS/COMBO-17; \citealt{ilb06}).  However the resolution attainable with the Hubble Space Telescope enables one to move beyond this to study multivariate distributions, such as the luminosity-size plane \citep{she03,dri05}, and the colour-structure plane \citep{dri06}.  These provide significantly more stringent constraints and are also now becoming the focus of numerical simulations (e.g., \citealt{alm06}). In this  article we summarise work under way \citep{cam07b} to investigate both of these multivariate distributions through the comparison of galaxies drawn from the Ultra Deep Field to our zero redshift reference sample defined by the MGC.  We pay particular attention to the multiple selection biases inherent in this work.

\section{Data}
\begin{figure}
\center
\includegraphics[width=58mm]{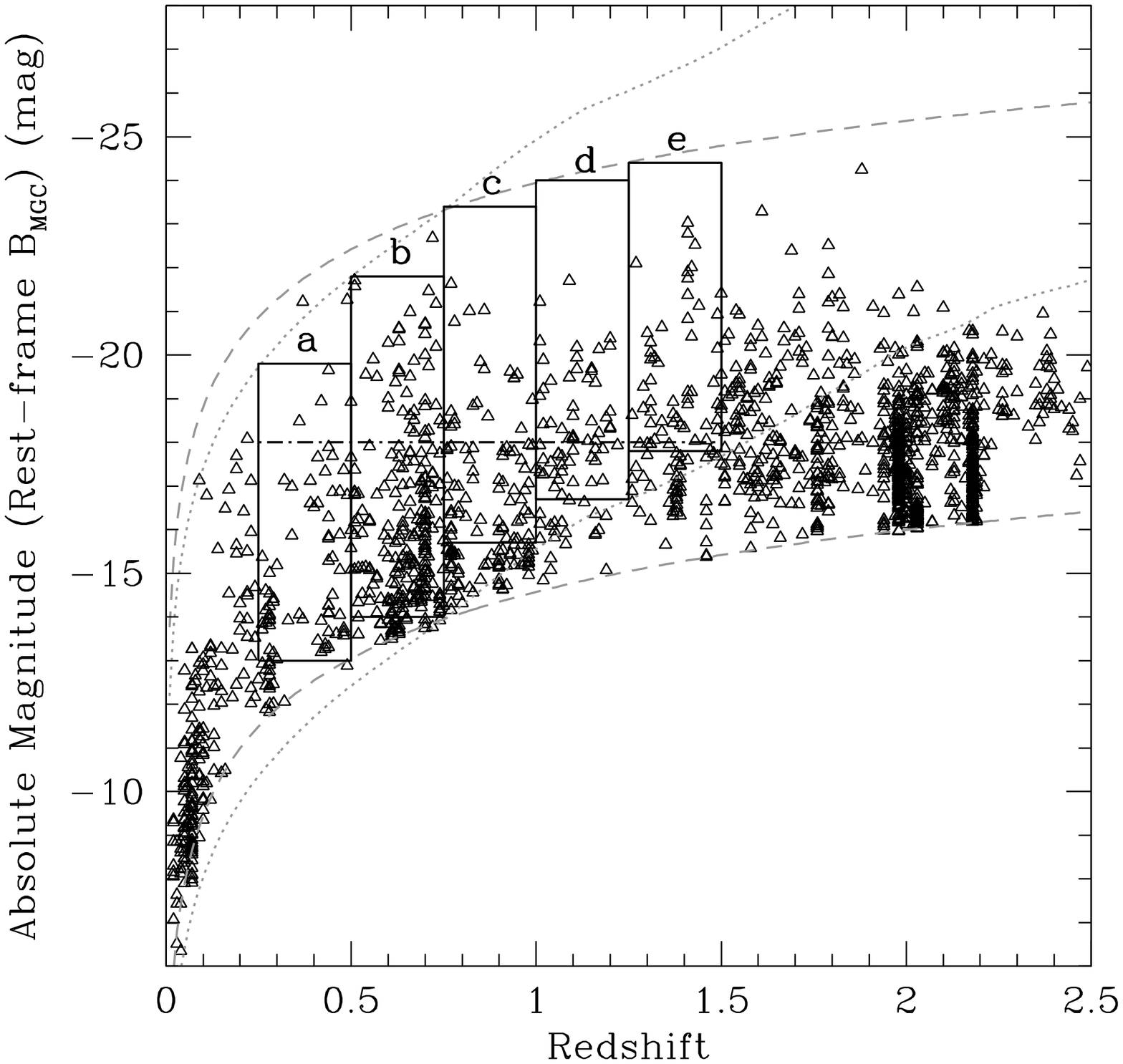}
\includegraphics[width=58mm]{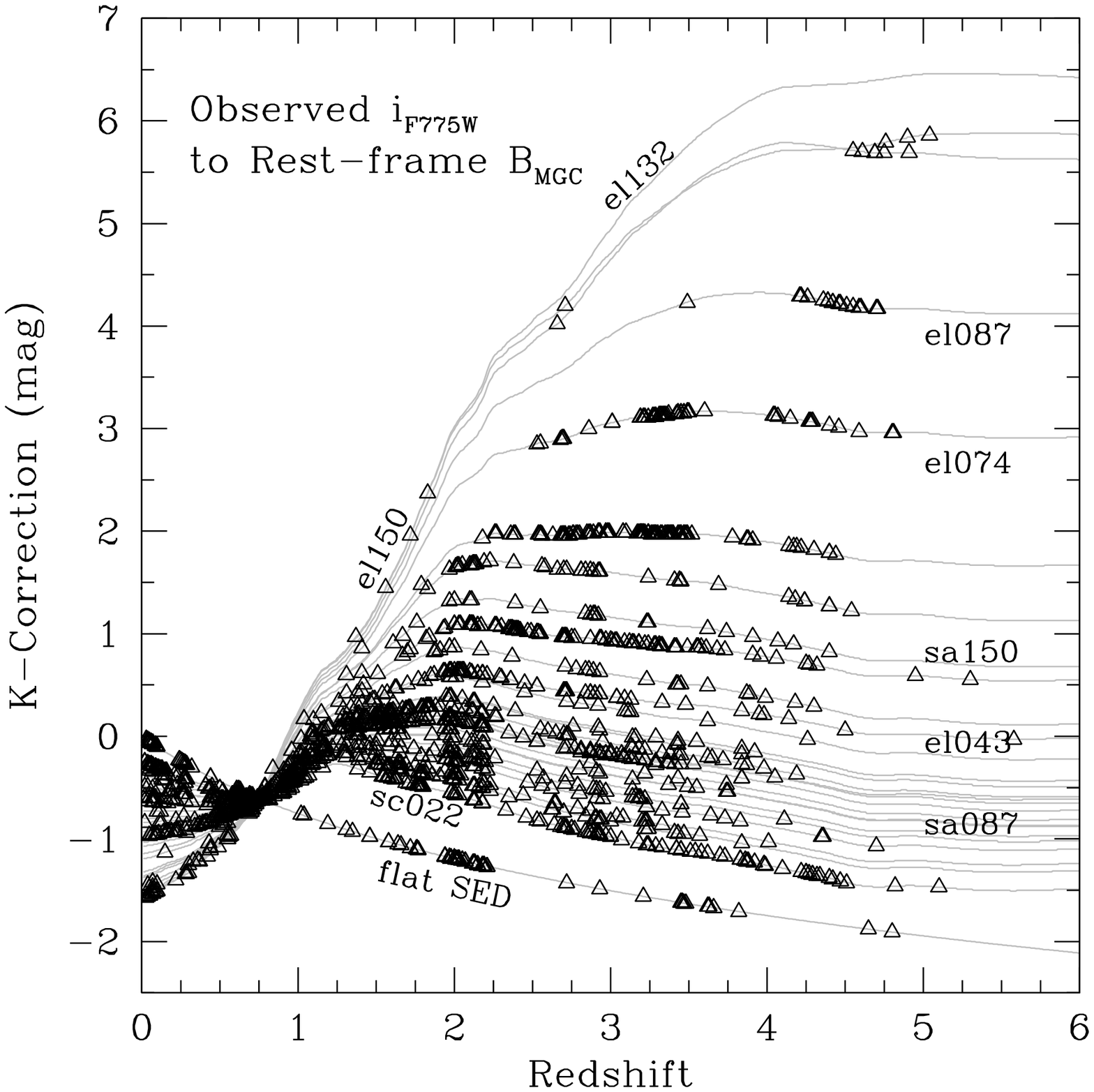}
\caption{(\textit{Left}) The iUDF-BRIGHT galaxy absolute magnitude - redshift distribution to $z = 2.5$. (\textit{Right}) K-corrections from observed ACS $i$-band to rest-frame MGC $B$-band for iUDF-BRIGHT galaxies to $z=6$.}\label{cameron_fig1}
\end{figure}

We use the Hubble Ultra Deep Field (see \citealt{bec06}) to sample galaxies out to $z=1.5$ and the MGC (see \citealt{lis03,dri05}) for a $z=0$ reference sample.  In \citet{cam07a} we present iUDF-BRIGHT, a sample of 2497 sources brighter than 28th magnitude in the ACS UDF $i$-band image.  These were detected by running SExtractor with a constant surface brightness threshold of 27.395 mag arcsec$^{-2}$.  Kron magnitudes and elliptical aperture half light radii were computed for all these objects.  Photometric redshift estimates based on $B$ through $H$-band aperture-matched magnitudes and global S\'ersic indices for 2172 galaxies in iUDF-BRIGHT were obtained from the catalogues of \citet{coe06}.  These redshifts and broadband magnitudes were used to derive individual galaxy K-corrections (observed ACS $i$-band to rest-frame MGC $B$-band) via fitting of SED templates from the library of \citet{pog97}.  We also included a flat spectrum model to approximate the K-corrections relevant to the extreme blue galaxies found in this type of deep imaging.  The iUDF-BRIGHT K-corrections are shown in Fig.\ \ref{cameron_fig1} (right) along with the relevant SED template tracks.  The gap between the reddest and bluest K-corrections becomes exceptionally large towards high redshift ($\sim$8 mag by $z \sim 4$), which means galaxies with these different SED types are observable over vastly different volumes at fixed luminosity.  This is a large potential source of selection bias if it is not accounted for in the subsequent analysis, and is one of the major drawbacks of optical wavelength studies at high redshift.\\

Using our K-corrections, the \citet{coe06} photometric redshifts and a cosmological model with $\Omega_{M} = 0.3$, $\Omega_{\Lambda} = 0.7$ and $H_{0} = 100$ km s$^{-1}$ Mpc$^{-1}$ we were able to compute absolute $B_{MGC}$-band magnitudes and sizes.  The iUDF-BRIGHT absolute magnitude - size distribution is shown in Fig.\ \ref{cameron_fig1} (left).  Broken, grey lines indicate the selection limits on bright and faint galaxies for our spectral types with the reddest and bluest K-corrections.  Rectangular boxes mark the bias free magnitude ranges of five volume limited samples selected over intervals of 0.25 in redshift from $z=0.25$ to 1.5.  The broken line indicates the $B_{MGC} < -18$ mag cut later implemented for exposing the colour-log($n$) relation.\\

\section{Selection Limits}
\begin{figure}
\center
\includegraphics[width=53mm]{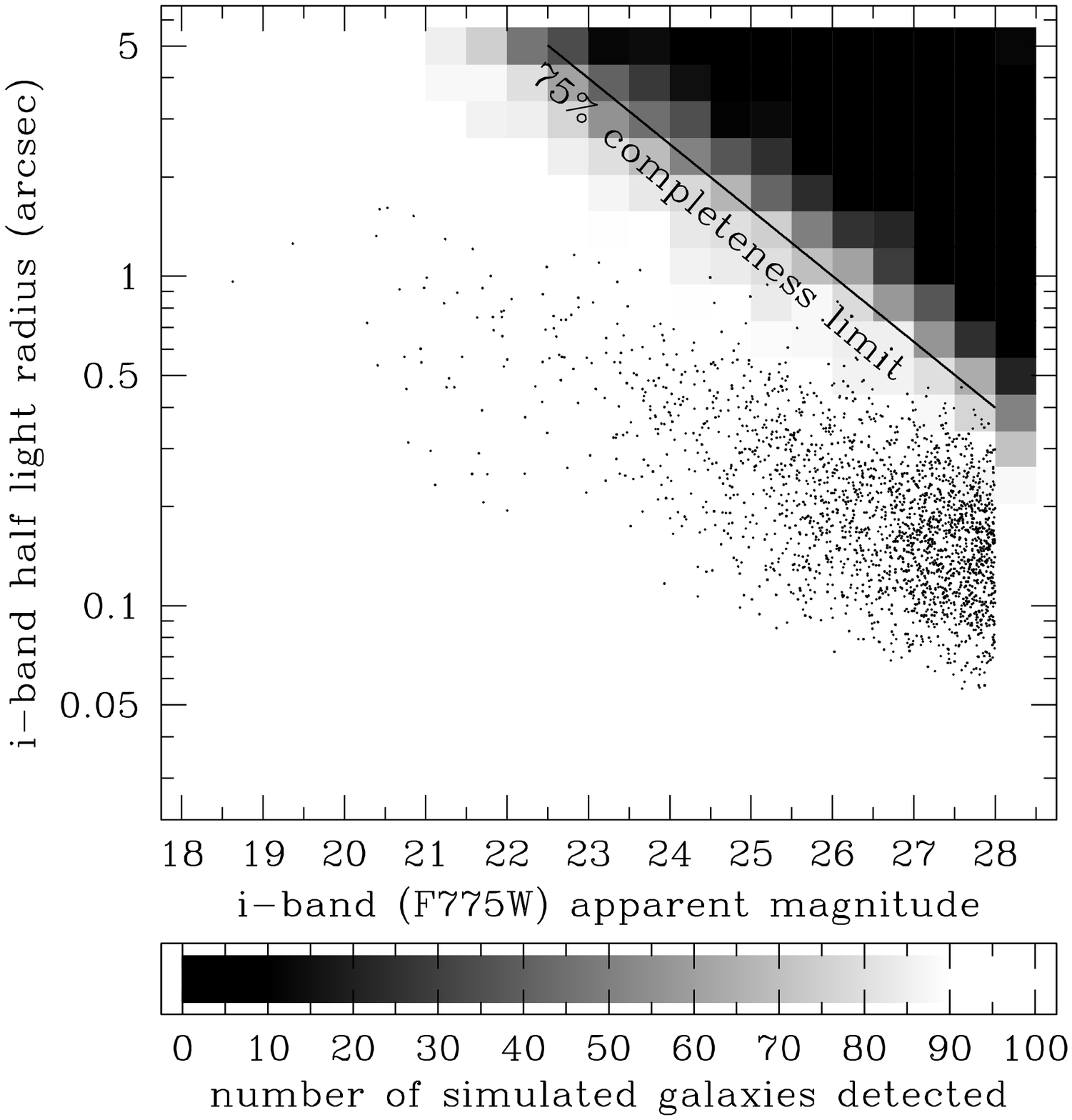}
\includegraphics[width=53mm]{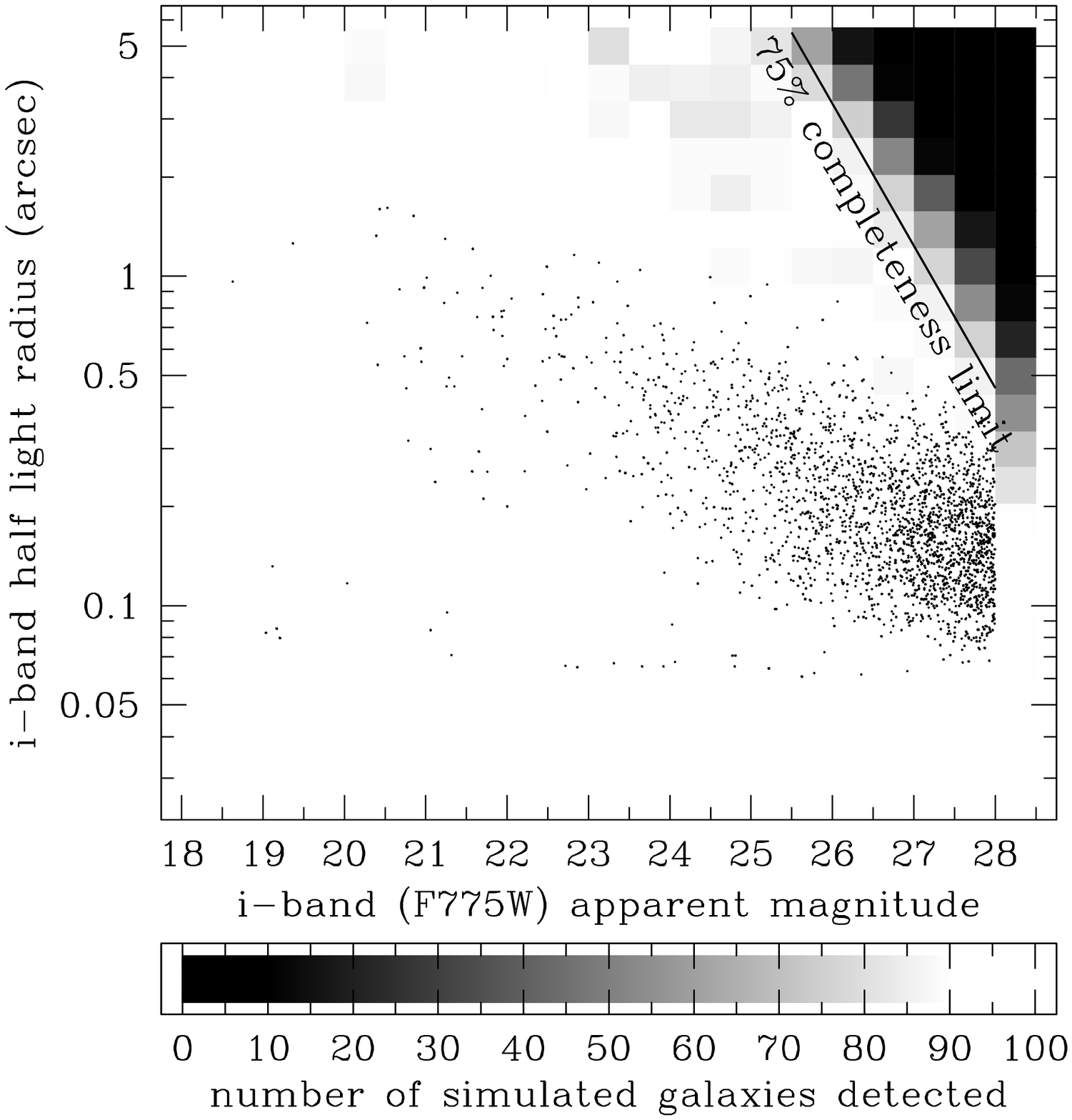}\\
\includegraphics[width=53mm]{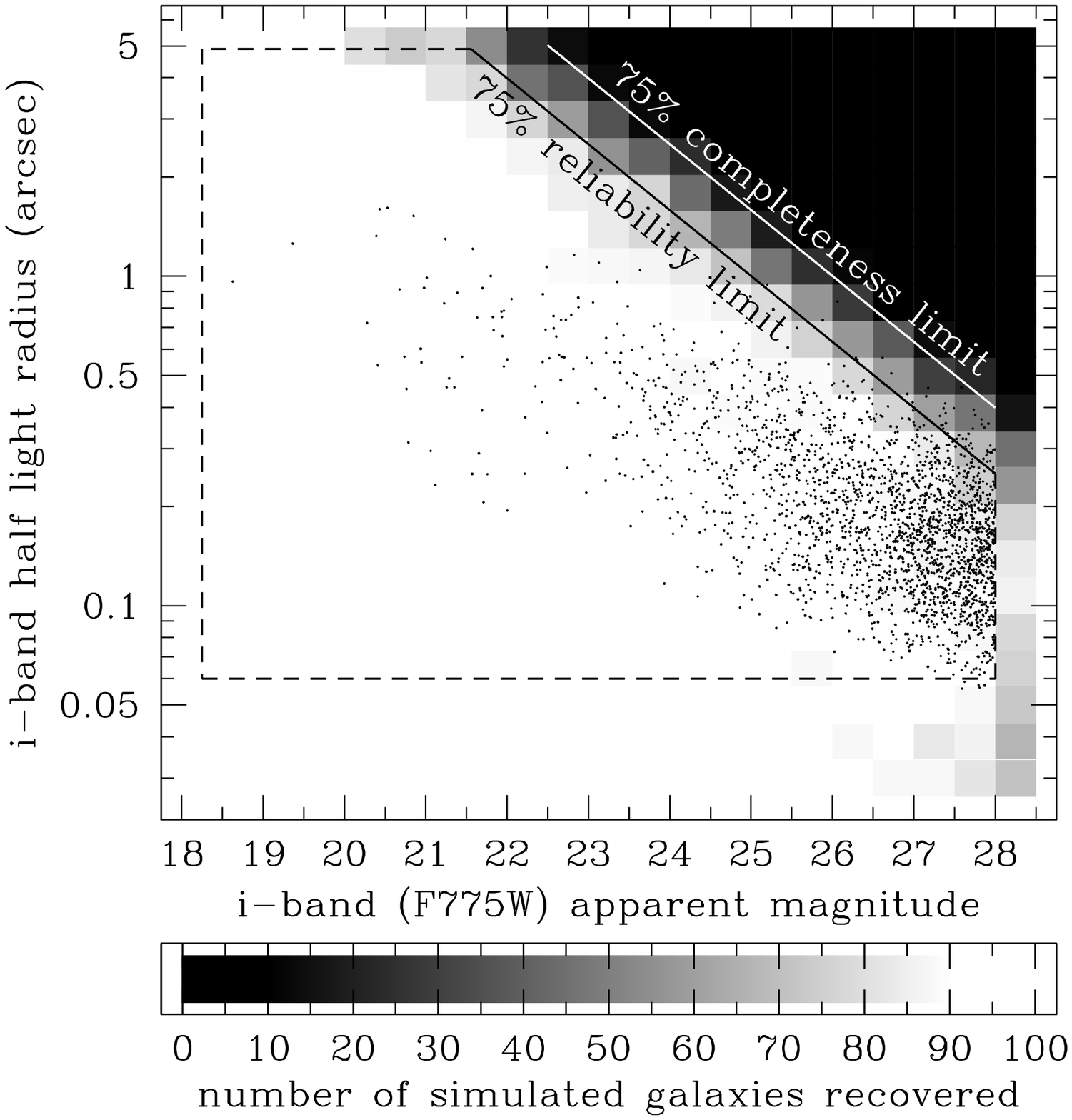}
\includegraphics[width=53mm]{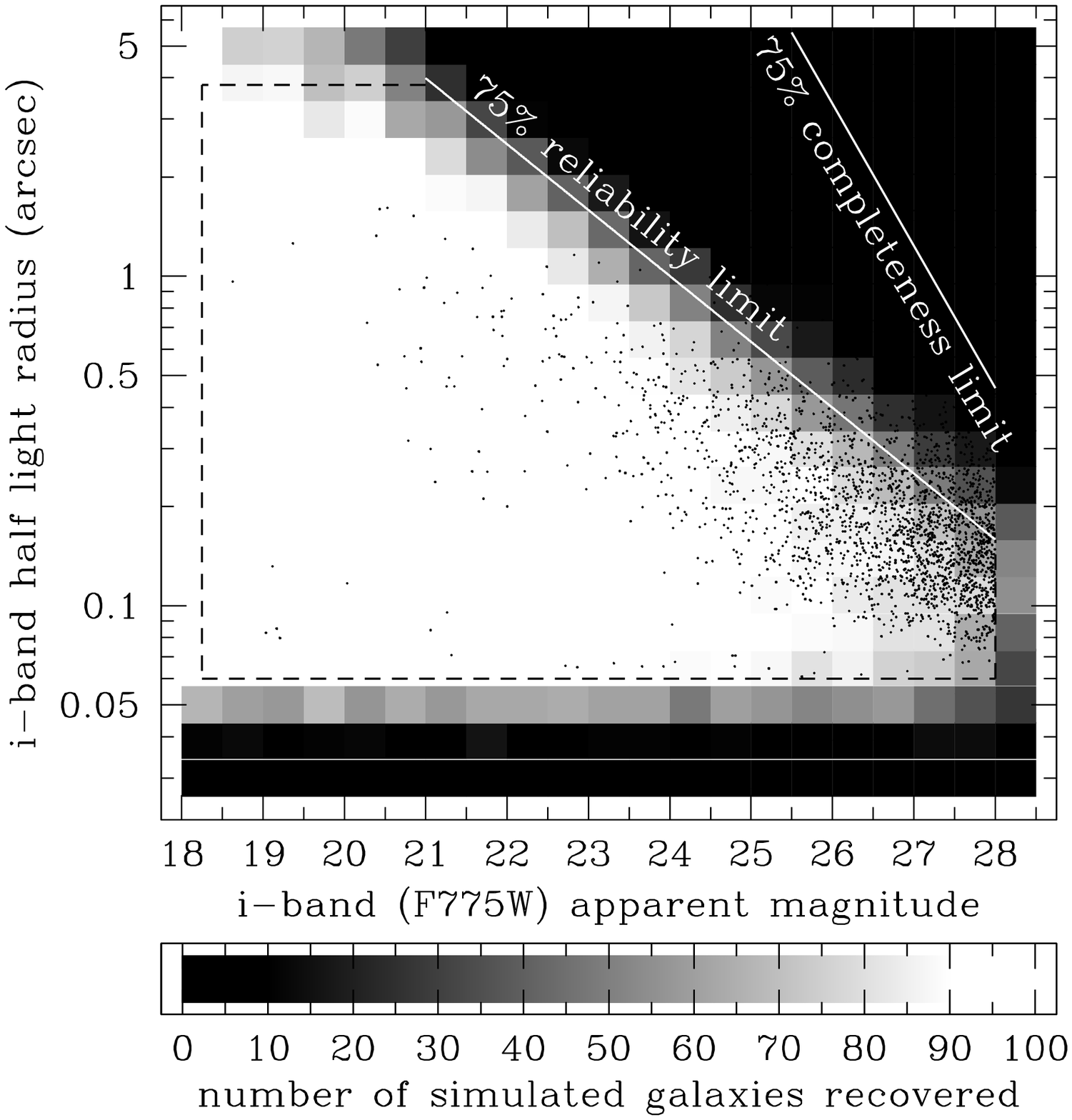}\\
\includegraphics[width=53mm]{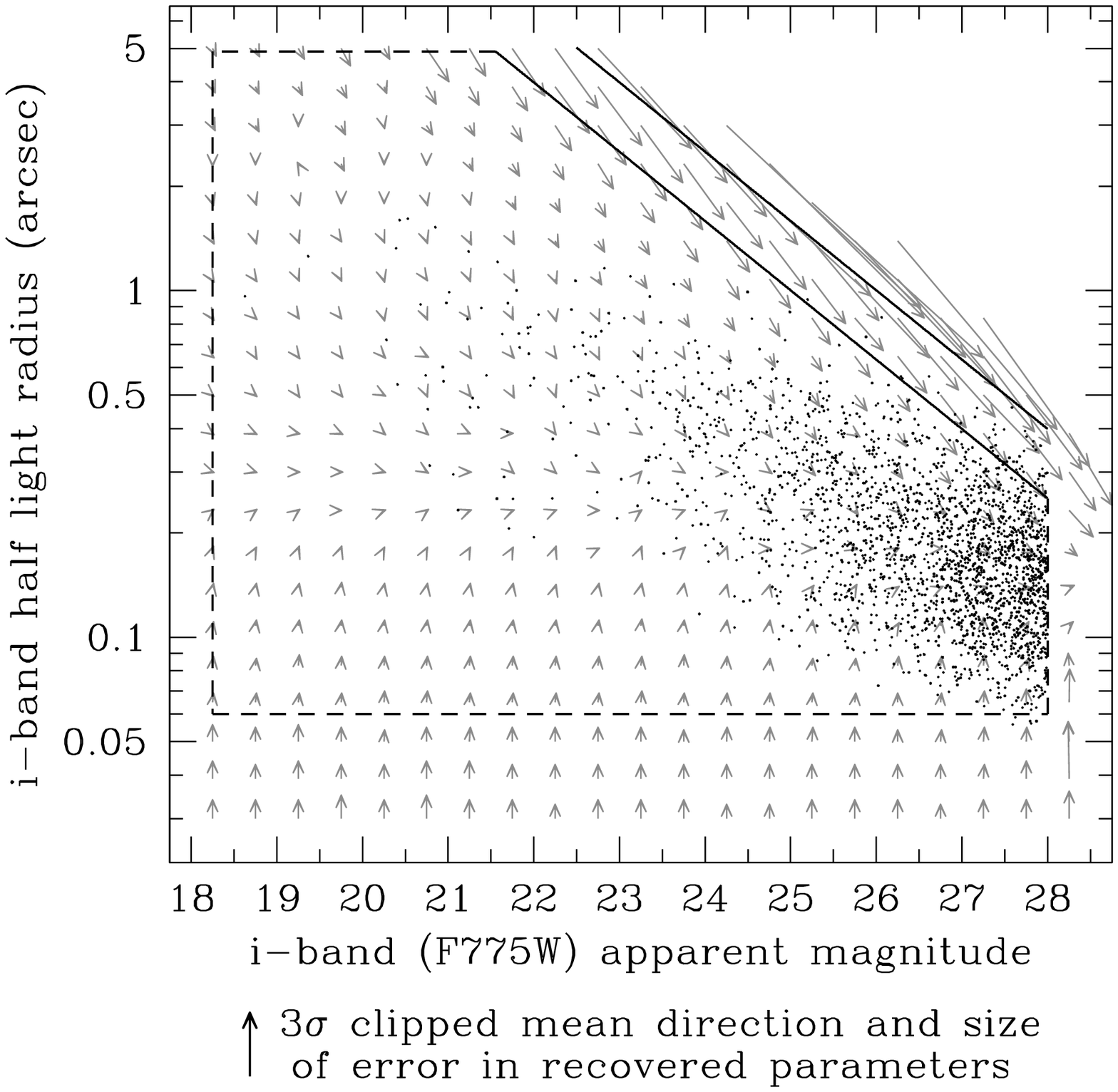}
\includegraphics[width=53mm]{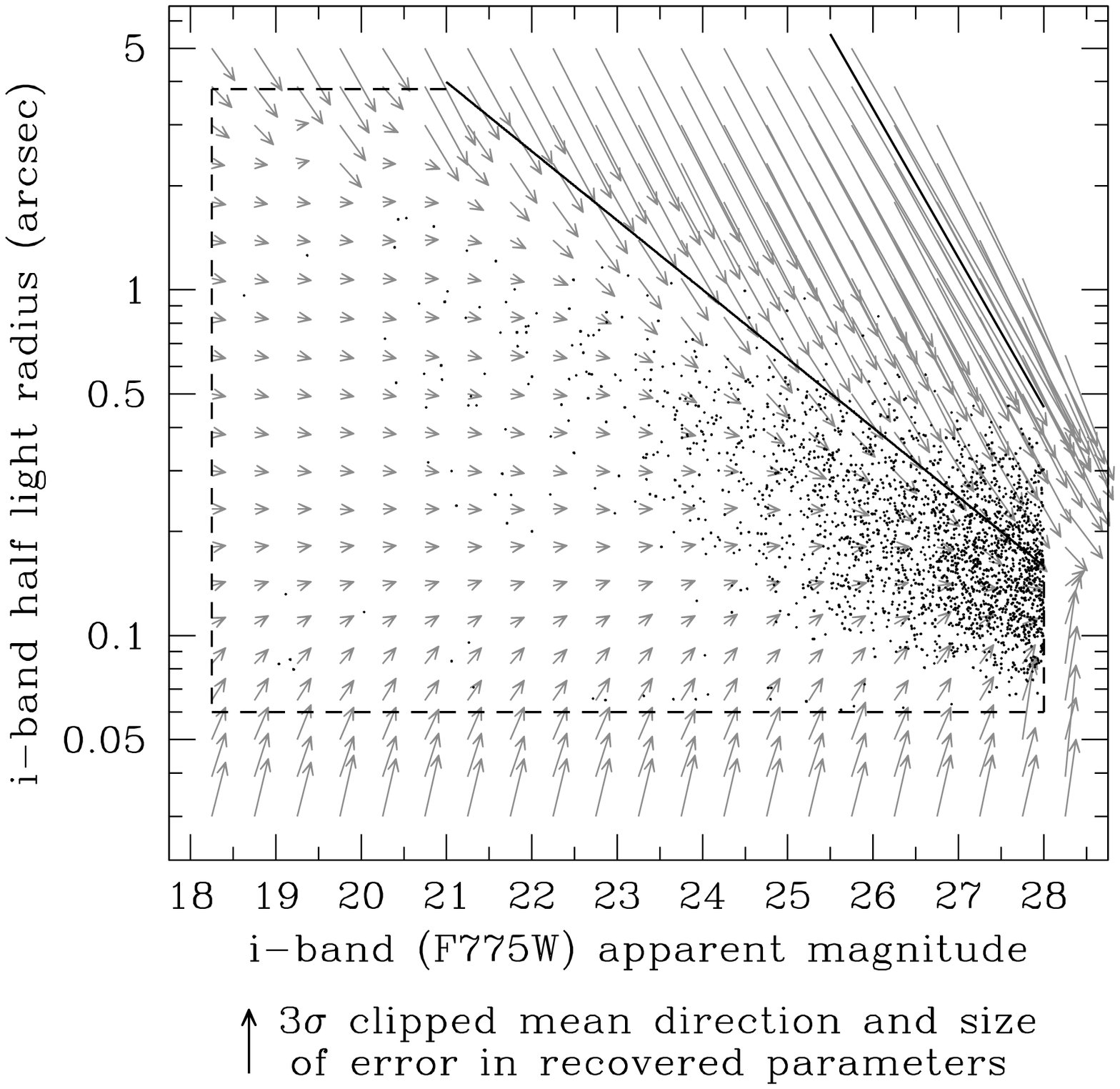}
\caption{(\textit{Top}) Completeness, (\textit{Middle}) reliability and (\textit{Bottom}) error vector diagrams for objects in the ACS UDF $i$-band image computed from our artificial galaxy simulations.  The left-hand column shows the results for an exponential (disk-like, $n=1$) profile and the right-hand column shows those for a de Vaucouleurs (elliptical-like, $n=4$) profile.  The observed iUDF-BRIGHT galaxy distribution is marked by the black dots.}\label{cameron_fig2}
\end{figure}

Artificial galaxy simulations are used to test our detection completeness and measurement reliability for galaxies in the ACS $i$-band image as a function of luminosity and size.  These simulations are described in detail in \citet{cam07a}.  In brief, we divide the luminosity-size plane into a grid and at each grid point generate 100 artificial galaxies using the IRAF \texttt{artdata} package.  These are inserted into the $i$-band image and passed through our detection and measurement pipeline.  The output detections, fluxes and sizes are compared to the input catalogue.  The results of these simulations for both exponential (diffuse, $n=1$, left-hand panels) and de Vaucouleurs (compact, $n=4$, right-hand panels) profile shapes are shown in Fig.\ \ref{cameron_fig2}.  Both types are readily detectable over the entire parameter space spanned by the observed data.  However, one cannot recover accurate fluxes and half light radii beyond mean surface brightnesses of $\sim$27.0 mag arcsec$^{-2}$ for the exponentials and only $\sim26.0$ mag arcsec$^{-2}$ for the de Vaucouleurs galaxies.  \textit{The error vector diagrams indicate these mis-measured galaxies are erroneously recovered as faint, compact systems!}  Although the de Vaucouleurs type galaxies are easier to detect (even at faint, mean surface brightnesses) due to their bright central cores, these profile shapes have extended low surface brightness wings, which contain a significant fraction of their total flux.  Hence, it is more difficult to \textit{accurately} recover their true magnitudes and sizes.  \textit{This illustrates the importance of considering reliability and profile shape when computing selection boundaries}.\\

\vspace{-0.5cm}
\section{The Colour-Log($n$) Bimodality}
\begin{figure}
\hspace{-0.95cm}
\includegraphics[width=146mm]{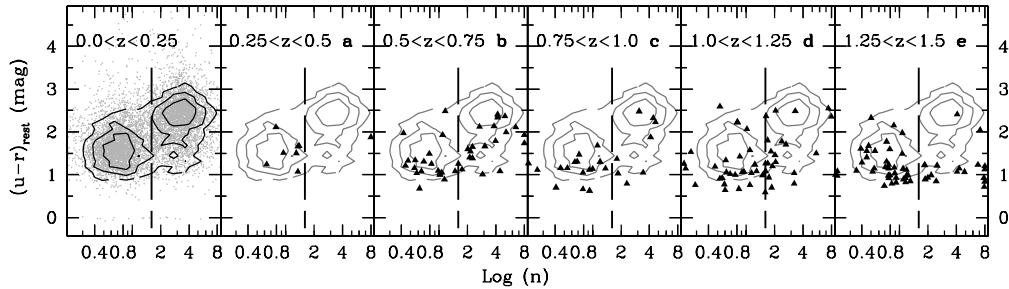}
\vspace{-10.5cm}
\caption{Colour-log($n$) diagrams for the MGC $0.0 < z < 0.25$ sample and our  five iUDF-BRIGHT volume limited samples from $z = 0.25$ to $z=1.5$.}\label{cameron_fig3}
\end{figure}

Using the broadband magnitudes and S\'ersic indices from \citet{coe06} we are able to construct colour-log($n$) diagrams for galaxies in each of our iUDF-BRIGHT volume-limited samples, as shown in Fig.\ \ref{cameron_fig3}.  We use our SED fits and K-correction software to correct from the observed filter combination straddling the 4000$\AA$ break at each redshift to rest-frame Sloan ($u$-$r$) colour.  \citet{dri06} have demonstrated that the local galaxy population shows a clear bimodality in the ($u$-$r$)$_{\mathrm{\tiny{rest}}}$-log($n$) plane for the Millennium Galaxy Catalogue sample.  This bimodality is most clearly seen for galaxies with abosolute magnitude greater than $\sim$$-18.0$ mag in the $B_{MGC}$-band.  If fainter galaxies are included, these are predominately found to be blue, diffuse systems, and the blue peak overwhelms the distribution.  Likewise, if only very bright galaxies are included, the bimodality is dominated by compact, red systems.  Even with this cut, our UDF volume-limited samples do not show the colour-log($n$) bimodality as clearly as the MGC data.  This is probably due in part to small number statistics and in part to the large errors in UDF magnitudes, colours and photometric redshifts.  In addition, there appears to be a certain degree of colour evolution towards bluer ($u$-$r$)$_{\mathrm{\tiny{rest}}}$ values at higher redshift, particularly for the more compact galaxies.  The dashed line indicates our cut at $n=1.6$ used to separate our sample into compact and diffuse galaxy types.  The $n=1.6$ cut was chosen to lie at the saddle point between the two MGC peaks.  We use a non-colour dependent divide as we expect a galaxy's S\'ersic index (i.e., its global structure) generally evolves on a longer timescale than its colour (i.e., its stellar population and recent star-formation history).\\

\vspace{-0.5cm}
\section{Luminosity-Size Diagrams}
\begin{figure}
\hspace{-0.53cm}
\includegraphics[width=141mm]{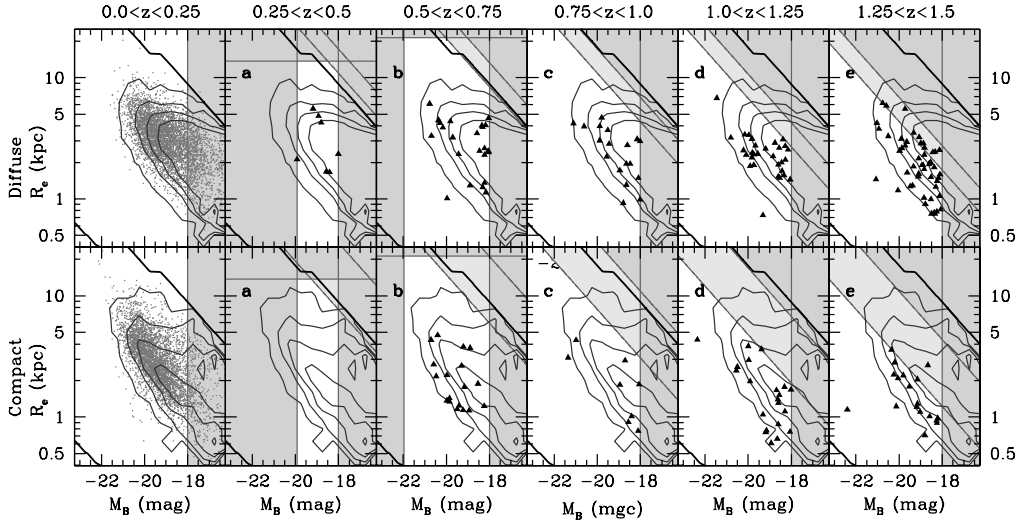}
\vspace{-7.25cm}
\caption{Luminosity-size diagrams for galaxies with $B_{MGC} < -18$ mag and $0.25 < z < 1.5$ in iUDF-BRIGHT, subdivided into diffuse ($n < 1.6$) and compact ($n>1.6$) structural types.} \label{cameron_fig4}
\end{figure}

In Fig.\ \ref{cameron_fig4} we present luminosity-size diagrams for galaxies in our volume limited samples with $B_{MGC} < -18$ mag and $0.25 < z < 1.5$, subdivided into diffuse ($n < 1.6$) and compact ($n>1.6$) structural types.  Regions of parameter space excluded by our selection limits are shaded in grey.  These account for detection completeness, measurement reliability and the K-corrections for red and blue galaxies.  They were constructed in the $M$-$R_e$ plane by combining our artificial galaxy simulation results with the method of \citet{dri99}.  The distribution of galaxies occupying the remaining white-space should be free of these biases.  The MGC local galaxy relation, also subdivided by structural type, is shown in the top panels and subsequently overlaid as contours.  In the diffuse galaxy samples \textbf{b}, \textbf{c} and \textbf{d} where the distribution is free of bias, there is evidence of evolution towards higher surface brightnesses at high redshift and perhaps a change of slope in the luminosity-size relation.  The compact galaxy samples are more tightly bound by the selection limits, but appear to show minimal evolution over this redshift range.\\

\vspace{-0.5cm}
\section{Discussion}
The trends seen in our luminosity-size and colour-log(n) diagrams suggest that bright, diffuse (i.e., disc) galaxies are still evolving structurally from high to low redshift.  The sense of this evolution is that the brightest disc galaxies are becoming larger towards low redshift.  This is consistent with the picture from inside-out models of galaxy formation such as \citet{tru05} who find $\sim$25\% growth of disk galaxies since $z \sim 1$.  Bright, compact (i.e., elliptical galaxies), on the other hand, seem to have finished their structural evolution at high redshift, but continue to evolve in colour to low redshifts.  This picture is broadly consistent with the hierarchical clustering scenario in which the brightest elliptical galaxies are assembled at intermediate-high redshift and their stellar populations passively evolve thereafter.  However, it is also compatible with a scenario in which spheroids form early via collapse and disc grow later (see \citealt{dri06}).  The evolution of the disc galaxy $M$-$R_e$ slope suggests we are seeing more than simple passive fading and secular evolution of the disc galaxy population.  For a more detailed investigation of these issues see \citet{cam07b}\\

\vspace{-0.75cm}
\begin{thebibliography}{}
\bibitem[\protect\citeauthoryear{Almeida, Baugh and Lacey}{2006}]{alm06} Almeida C., Baugh C. M., Lacey C. G., 2006, astro-ph/0608544
\bibitem[\protect\citeauthoryear{Beckwith et al.}{2006}]{bec06} Beckwith S. V. W., Stiavelli M., Koekemoer A. M. et al., 2006, ApJ, 132, 5, 1729-1755
\bibitem[\protect\citeauthoryear{Cameron \& Driver}{2007a}]{cam07a} Cameron E., Driver S. P., 2007, submitted to MNRAS
\bibitem[\protect\citeauthoryear{Cameron \& Driver}{2007b}]{cam07b} Cameron E., Driver S. P., 2007, in prep.
\bibitem[\protect\citeauthoryear{Coe et al.}{2006}]{coe06} Coe D., Benitez N., Sanchez S. F., Jee M., Bouwens R., Ford H., 2006, AA, 132, 926-959
\bibitem[\protect\citeauthoryear{Driver}{1999}]{dri99} Driver S. P., 1999, ApJ, 526, L69-L72
\bibitem[\protect\citeauthoryear{Driver et al.}{2005}]{dri05} Driver S. P., Liske J., Cross N. J. G. et al., 2005, MNRAS, 360, 81
\bibitem[\protect\citeauthoryear{Driver et al.}{2006}]{dri06} Driver S. P., Allen P. D., Graham A. W. et al., 2006, MNRAS, 368, 414-434 
\bibitem[\protect\citeauthoryear{Ilbert et al.}{2006}]{ilb06} Ilbert O., Lauger S., Tresse L. et al., 2006, A\&A, 453, 3, 809-815
\bibitem[\protect\citeauthoryear{Lilly et al.}{1995}]{lil95} Lilly S. J., Tresse L., Hammer F. et al., 1995, ApJ, 455, 108
\bibitem[\protect\citeauthoryear{Liske et al.}{2003}]{lis03} Liske J., Lemon D. J., Driver S. P. et al., 2003, MNRAS, 344, 307L
\bibitem[\protect\citeauthoryear{Poggianti}{1997}]{pog97} Poggianti B. M., 1997, A\&A Supp., 122, 399-407
\bibitem[\protect\citeauthoryear{Shen et al.}{2003}]{she03} Shen S., Mo H. J., White S. D. M. et al., 2003, MNRAS, 342, 978-994
\bibitem[\protect\citeauthoryear{Thompson et al.}{2005}]{tho05} Thompson R. I., Illingworth G., Bouwens R. et al., 2005, ApJ, 130, 1, 1-12
\bibitem[\protect\citeauthoryear{Trujillo and Pohlen}{2005}]{tru05} Trujillo I., Pohlen M., 2005, ApJ, 630, L17-L20
\end{thebibliography}
\end{document}